# Heat and Economic Preferences

Michelle Escobar Carias[1], David W. Johnston, Rachel Knott, Rohan Sweeney

Centre for Health Economics, Monash University

*This version: August 29th, 2022*

**Abstract**

Does temperature affect economic preferences and the likelihood of irrational behavior? Using data from Indonesia, we answer this question by estimating how risk aversion, impatience and the occurrence of a rational choice violation vary with outdoor temperatures within areas over time. Our findings show that hot weather causes more rational choice violations and increases impatience, but does not affect risk-aversion. These effects are driven by temperatures on the night prior to the survey rather than by temperatures during the day of the survey. An important mechanism behind these effects is depleted cognitive functioning, particularly mathematical skills. These findings suggest that heat-induced night-time disturbances cause stress on the brain, which then manifest in significantly lower cognitive functions critical for economically rational and utility-maximizing decision-making. This has important implications for poorer households in low- and middle-income countries, who are most exposed to extreme heat.

**Acknowledgements:** The authors are grateful to Maximilian Auffhammer, Nathan Kettlewell, Nicole Black, Claryn Kung, Umair Khalil, seminar and conference participants at the Nordic Conference in Development Economics, UCLA & Luskin Climate Adaptation Research Symposium, CEPR Women in Economics Workshop, Symposium on Economic Experiments in Developing Countries, Applied Young Economists Webinar, Monash Environmental Economics Workshop, Australasian Development Economics Workshop and the Asian and Australasian Society of Labour Economics Conference for useful comments and suggestions. All errors are our own.

---

[1] Corresponding author: michelle.escobarcarias@monash.edu. Address: 900 Dandenong Road, Centre for Health Economics Monash University, Caulfield East, 3145, Victoria, Australia.

# 1. Introduction

People's economic preferences influence types of behavior that are important for individual economic well-being as well as for broader social welfare. For instance, studies show that savings behavior and investments in human capital inputs, such as child immunizations and schooling, depend partly on a person's level of patience (Ashraf et al., 2006; Banerjee et al., 2010). Similarly, risky choices linked to economic development, such as opening a new business, becoming self-employed, and the optimal use of fertilizer, depend on a person's willingness to accept risks (Ekelund et al., 2005; Liu & Huang, 2013). In the neoclassical theory, these types of individual preferences are assumed to be stable across time (Stigler & Becker, 1977). However, a growing body of evidence shows that preferences change after exposure to negative shocks, such as natural disasters, violence, bereavement and financial crises, with the size and direction of effects varying substantially across studies (Cameron & Shah, 2015; Hanaoka et al., 2018; Callen et al., 2014; Callen, 2015; Kettlewell, 2019; Levin & Vidart, 2020).

In this study, we investigate whether economic preferences and rational behavior are impacted by temperature. According to the latest report from the Intergovernmental Panel on Climate Change (IPCC, 2021), human-induced climate change has already increased the level of average temperatures and the frequency of temperature extremes over land. Evidence suggests that these temperature spikes could lead to significant changes in decision making due to their negative impacts on cognition, sleep patterns and mood. For instance, Mani et al. (2013) and Schilbach et al. (2016) show that mentally taxed individuals are less likely to engage in slower, deliberative and logical thinking – what Kahneman (2011) refers to as System 2 thought processes – and these are required for making complex calculations and decisions. Schilbach et al. (2016) propose that individuals have a limited System 2 budget, and factors such as sleep deprivation and heat can reduce a person's limited mental bandwidth. This hypothesis has been broadly corroborated by a number of economic studies. Examples include high-stakes decisions by judges (Heyes & Saberian, 2019), consumer behavior and purchase decisions (Busse et al., 2015), violent interactions among incarcerated populations (Mukherjee & Sanders, 2021) and even criminal behavior in poor neighbourhoods of the United States (Heilmann et al., 2021).

We use data from Indonesia to estimate the effects of temperature on risk aversion, impatience and rational choice violations.[2] Weather data come from NASA's Modern-Era Retrospective Analysis for Research and Applications (MERRA-2). Elicited rational choice, and time and risk preference data come from the Indonesia Family Life Survey (IFLS). To identify causal effects, we use short-term variations in temperature over time within climate zones defined by the interaction of a village's province, altitude, distance to the coast, and level of urbanity. Subsequently, we estimate the effects of temperature on cognition, sleep, and mood in order to better understand the mechanisms driving the relationships between temperature and economic preferences. These three outcomes have been shown to be negatively impacted by extreme heat (Zivin et al., 2018; Garg et al., 2020; Park et al., 2020a; 2020b; Minor et al., 2020; Obradovich et al., 2017; Mukherjee & Sanders, 2021; Almås et al., 2020; Baylis, 2020) and to be associated with decision making (Benjamin et al.,2013; Nofsinger & Shank, 2019; Cueva et al., 2015).

Our findings show that higher temperatures lead to significantly increased rational choice violations and higher impatience, but do not significantly affect risk aversion. We show that these effects are mainly driven by temperatures on the night prior to the survey rather than by temperatures on the day of the survey. The estimates indicate that a 1°C increase in midnight temperature is associated with a 1 percentage point (3.2%) increase in irrational behavior and a 0.9 percentage point (1.5%) increase in impatience. Our exploration of potential mechanisms indicates that higher night-time temperatures significantly reduce cognitive functioning the following day - mathematical skills in particular - at a rate of 1.6%-3.6% of a standard deviation for each degree Celsius. However, we detected no effects on self-reported sleep duration or mood. Based on these findings, we posit that heat-induced night-time disturbances cause stress on parts of the brain, such as the prefrontal cortex (Miller, 2000), which manifest in significantly lower cognitive functions the next day. These functions are important for individuals to perform economically rational decision-making, and therefore high night-time heat leads to an increase in sup-optimal decision-making.

We additionally explore non-linear temperature effects and heterogeneous treatment effects by individual and household characteristics. The former exploration shows a linear increase in rational choice violations and impatience when midnight outdoor temperatures are above 22°C. Given that 24% of our sample experienced midnight temperatures above this threshold the night prior to their survey, these findings are relevant for a large portion of the Indonesian population. With regards to heterogeneity, we find that temperature effects are larger for individuals with lower schooling, those who reside in households with lower asset wealth, and in households with low electricity expenditure. These results suggest that poorer households are less able to mitigate their exposure to extreme outdoor temperatures, and therefore are the most negatively affected by heat in terms of the quality of their decision making.

---

[2] Our measure of rational choice violation follows from individuals entirely avoiding entering a more than fair lottery. This could also be considered a non-standard preference such as individuals having a certainty premium (Callen et al., 2014) and/or regret aversion (Zeelenberg et al., 1996), which are not necessarily considered irrational. However, in our sample, cognitive abilities are highly and negatively correlated with making 'rational choice violations' but not with risk aversion, an actual preference. We therefore argue that this is more likely than not, evidence of rationality errors from which individuals obtain no positive utility.

These findings contribute to a small collection of studies that investigate the effect of temperature on time and risk preferences in laboratories. Overall, these studies show little consensus. Almås, et al. (2020) studied the impact of heat on university students in Berkley, California, and Nairobi, Kenya. They find no evidence that heat stress affects risk-taking, time preferences, cognition, or social behavior except for destructive tendencies.[3] In contrast, Wang (2017) finds that high ambient temperatures lead individuals to pursue high-risk options. A third study conducted by Cheema & Patrick (2012) finds that individuals exposed to warmer temperatures in the lab are less likely to gamble or purchase innovative products.

Our study contributes to this literature in several ways. Although the temperature variations we use are not randomized as they are in laboratory studies, we leverage a quasi-experimental design in which temperatures are as-good-as-randomly assigned (an assumption we extensively investigate). Further, we overcome some general limitations of laboratory experiments, such as small sample sizes and limited experimental subject pool. Our sample of approximately 50,000 individuals ages 15 to 90 years old is significantly larger than the highest-powered laboratory study with 2,000 observations (Almås, et al., 2020). Further, our findings are more generalizable given our use of a (nearly) nationally representative sample. This helps increase the external validity of our results for populations residing in low- and middle-income countries. Finally, in contrast to the cited experimental studies, we do not assume an immediate impact of heat on preferences. Instead, our data allow us to identify heterogeneous responses to timing and duration of heat exposure.

We also contribute to the mixed literature on the relationship between economic preferences and extreme natural, social and economic events as well as to the literature on the effects of poverty on economic preferences (Bartoš et al., 2021; Carvalho et al., 2016). For example, Cameron & Shah (2015) find that Indonesians who have experienced a flood or an earthquake exhibit more risk-averse tendencies, while conversely, Hanaoka et al. (2018) find that victims of the 2011 Great East Japan Earthquake became more risk-tolerant and increased gambling. In other related literature, studies have found that environmental conditions can increase the likelihood of making violations of rational choice in purchasing decisions. For example, Busse et al. (2015) find that sales of convertibles and four-wheel drive vehicles are significantly influenced by idyosincratic shocks in temperature, cloud cover and snowfall. In a similar vein, Conlin et al. (2007) find that the probability of customers returning catalog orders of cold-weather items is significantly increased when order-date temperatures drop. Both are examples of how environmental conditions may increase the propensity for rational choice violations, which are costly to the individual and potentially to society.[4]

---

[3] Almås, et al. (2020) find that heat affects people's willingness to destroy other people's earnings in a game known as joy of destruction, consistent with the temperature-aggression hypothesis.

The rest of the paper proceeds as follows. Section 2 outlines the sources of data, outcomes and treatment measures. Section 3 outlines the empirical strategy. Section 4 provides a discussion of the main findings, robustness checks and placebo tests. Section 5 provides a discussion regarding potential mechanisms behind the identified temperature effects and section 6 concludes.

**2. Data**

*2.1 Indonesia Family Life Survey*

Data on time and risk preferences, rational choice, cognition, sleep, and mood come from waves 4 and 5 of the Indonesia Family Life Survey.[5] These two survey waves spanned 2007-2008 and 2014-2015, and include over 50,000 completed surveys from respondents residing in more than 3,800 villages across 38 provinces. All IFLS outcomes used in the analyses were collected verbally in face-to-face sessions with the survey respondents. Descriptions and summary statistics of the key IFLS variables are shown in Table 1, and are further detailed in the below subsections.

---

[4] For instance, according to The Guardian (2020), about five billion pounds of returned goods in the United States ended up in landfills in 2019, which contributed 15 million metric tons of carbon dioxide to the atmosphere.
[5] The time and risk preference modules were not collected in waves 1-3 of the Indonesia Family Life Survey.

*2.1.1 Risk and Time Preferences*

The IFLS risk aversion module is a staircase-type instrument that includes a series of hypothetical choices between accepting a certain amount of money or accepting a risky lottery that offers a 50% probability of winning an amount '$c$ times smaller' 'or $x$ times larger' than the certain value. The factors $c$ and $x$ are adjusted with subsequent tasks depending on previous choices (see Appendix Figure A1 for the complete sequence of potential choice tasks).

We use the first-choice task in the module to capture violations of expected utility (rational choice violations). In it, individuals are given a hypothetical choice between a certain amount or a 50% chance of winning that amount or twice that amount. The rational option is the coin flip, as per the non-satiation assumption (Perloff, 2014). Nevertheless, 32% chose the certain option and refused the gamble. We use a binary indicator of this choice as our measure of a rational choice violation.[6]

The remaining 68% of respondents choose the 'risky' option in the first-choice task and are given subsequent decisions that can be categorized into four groups: most risk averse (27.9%), second most risk averse (8.8%), third most risk averse (10.6%) and least risk averse (17.5%). The outcome variable in our regression analyses is a binary variable equalling one if the individual is in the most risk averse group, and zero if in one of the other three groups. By this definition, 37.3% of this sample is considered risk-averse.

Our measure of impatience is derived from a similar staircase-type instrument with a series of hypothetical choices beginning with earning a certain amount today or that same amount at a future date. The first value is held constant in each subsequent question, and the second value increases throughout the series to capture different discount rates (See Appendix Figure A2 for the complete sequence of potential choice tasks). We construct a binary measure for myopic preferences where respondents are assigned a value of one if their sequence of decisions categorises them in the most impatient group and zero otherwise. Approximately 62% of individuals are in the most impatient category.[7] All three main outcome variables are dichotomized to ease interpretation.

---

[6] It is important to note that this lottery choice sequence was not monetarily incentivized, a relevant feature of the risk and time modules because the majority of the Indonesian population identify themselves as Muslim and gambling is forbidden in Islam. Furthermore, to account for the possibility that our 'rational choice violation' outcome might partly capture reluctance to gamble due to religious beliefs, our main regression models control for religion and religiosity. Moreover, as shown in Appendix Table B1, individuals who identify as Muslim are not statistically more likely to make rational choice violations (-1.1 percentage points, $p = 0.846$).

We validate these economic preference measures by estimating the associations between them and individual behaviours. Following the approach used by Falk et al. (2018) to validate the measures obtained in the Global Preference Survey (GPS), we create two groups of behavioural outcomes – accumulation decisions and risky choices – and regress these outcomes on either impatience, risk aversion, or the probability of making rational choice violations.

Economic theory relates patience with the ability of individuals to save and invest in education in order to accumulate financial and human capital. If an individual's marginal utility of consumption in the present is very high or their expectations of the future are very low (i.e. myopic or impatient), she will be less inclined to save and invest. In columns 1 and 2 of Table 2 we show that indeed, for our sample of IFLS respondents, impatience[8] is associated with a 3.3-percentage point reduction in the probability of saving and an 8.8 percentage point reduction in the probability of achieving secondary education or higher.

Similarly, we investigate whether rational choice violations[9] and risk aversion are significantly associated with risky behaviour: self-employment, planning to open a business, and using tobacco. Our findings indicate that both risk-averse individuals and people who make rational choice violations are less likely to be employed and plan to start a business (columns 3 and 4, Table 2). Conditional on smoking, both consume a significantly lower number of cigarettes (columns 5 and 6, Table 2).

*2.1.2 Cognition, Sleep and Mood*

Four cognitive tests were administered to respondents age 15 to 90 in wave 5 of the IFLS: (i) an adaptive number series test measuring fluid intelligence related to quantitative reasoning (Strauss et al., 2017); (ii) a date awareness, word recall, and subtraction test measuring 'mental intactness' (Fong, et al., 2009); (iii) the Raven's Progressive Matrices test that measures non-verbal abilities (Raven, 2000); and (iv) a test of mathematical skills with addition, subtraction, multiplication and division questions (see the Supplementary Online Appendix for more details on each test). We standardized the score from each test such that it has a mean of zero and a standard deviation of one. We also use a standardized composite measure of cognition by performing a principal component analysis of the four tests. In wave 4 of the IFLS, some of these tests were only asked of adults aged 15-31 years, and one item – fluid intelligence – was not included at all.

---
[7] As with risk aversion, the first decision in the time preference task captures a non-standard preference, in this case, negative time discounting. Negative time discounters form only 2% of our sample and are subsequently excluded from the analysis.
[8] In our sample, males, people with a lower likelihood of having secondary schooling and people with lower cognitive scores are more likely to be impatient as shown in Appendix Table B1.
[9] In our sample, females, unemployed individuals, people without secondary schooling and those with low cognitive scores are significantly more likely to make rational choice violations. See Appendix Table B1 for further details.

Wave 5 of the IFLS also asks respondents the time they went to bed the night before the survey and the time they woke the morning of the survey. Using these two variables, we construct a proxy for sleep duration. On average, our sample of adults went to bed at around 10:30 pm, woke up at 5:12 am, and spent approximately 6.8 hours in bed the night before the survey. In the absence of objective sleep measurements, we use this variable, but acknowledge its limitations. It is known that self-reported sleep duration is usually greater than objectively measured sleep duration (Schokman et al., 2018; Lauderdale et al., 2008). Also, time spent asleep fails to capture sleep quality or sleep efficiency. For example, Bessone et al. (2021) find in a study of low-income adults in Chennai, India that although on average their sample spend 8 hours in bed, they effectively sleep only 5.5 hours a night.

To test whether we can find a similar effect of temperature on mood as reported by Almås et al. (2020), we extract four questions from the IFLS5's Positive and Negative Affects module. This module asks individuals how angry, tired, enthusiastic or happy they felt 'yesterday', the day before the survey. As explained in Table 1, the responses 'quite a bit' and 'very' are aggregated to create binary variables. On average 31% of adults reported having felt angry, 45% felt tired, 58% felt enthusiastic, and 64% felt happy on the day before the survey.

*2.2 NASA MERRA-2 Dataset*

We obtain hourly and daily records of temperature, relative humidity, precipitation, wind speed, and pollution from NASA's MERRA-2 dataset. This dataset integrates both station and satellite data to provide localized indicators across the extensive landmass of the Indonesian territory, including areas with no proximal weather stations. Specifically, MERRA-2 provides environmental estimates for 0.5º x 0.625º cells (approximately 50 kms x 60 kms) at hourly and daily time scales from 1981 to 2021 (Rienecker, et al., 2011).[10]

---

[10] Auffhammer et al. (2013) caution that the output for regions where stations are sparse or of poor quality is a model prediction, and therefore likely to be less accurate than for regions with greater coverage of weather

MERRA-2 climate records are matched to IFLS respondents through the GPS coordinates of their villages of residence. Village level GPS coordinates were only collected in wave 1 (1993) and wave 5 (2014-2015) and so approximately 30% of wave 4 respondents (2007-2008) were omitted from the sample because they had no GPS coordinate data.[11] An additional 4.5% of respondents were omitted because their village was located at a distance greater than 50 kilometers from the closest MERRA-2 grid point. This restriction was made to reduce measurement error in the temperature variable. Figure 1 presents a map of the Indonesian territory with markers for the MERRA-2 grid points and each Indonesian village in wave 5 of the IFLS.

We use three temperature variables in the regression analyses: temperature at the beginning of the survey, maximum temperature on the survey day, and midnight temperature on the night prior to the survey. The mean maximum temperature on survey dates was 28.7ºC, with 1.4% of individuals experiencing a maximum temperature above 35ºC. The mean temperature in the first hour of the interview was 24.6ºC, and the mean midnight temperature on the night prior was 23.6ºC.

## 3. Empirical Strategy

*3.1 Defining Climate Zones*

We identify the effects of temperature on time and risk preferences by exploiting differences in temperature across time within homogenous climate zones. A key identifying assumption of this approach is that temperature variation within climate zones is exogenous. That is, we assume that individuals within the same climate zone face different temperatures only because of differences in survey dates.[12] This assumption is violated if there are significant differences in climatic conditions within the same area. For example, people living at sea-level may experience systematically hotter temperatures than those living at higher altitudes, and may also experience differences in socioeconomic and cultural environments that are associated with economic preferences. For this reason, rather than comparing differences between people

---

stations. This is especially relevant when using fixed effects estimators where it can amplify statistical concerns such as attenuation bias.

[11] The majority of wave 4 respondents had GPS coordinates because they were residing in a village visited by surveyors in wave 1 or wave 5. Omitted wave 4 respondents were those living in a 'new' village.

[12] Appendix Figure A3 shows the number of days it took to roll out the survey instrument in each climate zone with an average of 40 days.

living within politically-defined areas that often encompass different climatic conditions, we define our own areas.

Three key determinants of temperature within regions of a country are altitude, distance to the sea, and the level of urbanity (NASA, 2016; NOAA, 2021; EPA, 2021). Altitude, the height measured from sea level, is inversely correlated with temperature (Pielke, 2021). Costal temperatures tend to be cooler than inland temperatures, and thus distance from the sea is positively correlated with temperature. Finally, cities can cause an 'urban heat island effect', which occurs when cities replace natural land cover with dense concentrations of pavement, buildings, and other surfaces that absorb and retain heat (EPA, 2021). We therefore compare survey respondents living in the same Indonesian province and with similar altitude, distance to the sea, and level of urbanity. Specially, we create 285 climate zones by interacting: 24 province indicators, four altitude groups (<50 meters, 50-100 meters, 100-500 meters, 500+ meters)[13], three distance-to-the-coast groups (<30 km, 30-60 kms, 60+ kms), and an indicator of the village's urban or rural status.[14] The map in Figure 2 provides an example of the 19 resulting climate zones in the province of North Sumatra. Elevation is drawn in the background as shown by the grey shading, with darker areas representing higher altitude zones. The different climate zones used for the area fixed effects models are represented by different colours located at the centre point of each surveyed village. Our identification strategy implies, for instance, that we compare preferences across time of people residing in rural mountainous villages far from the coast (blue circles).

Figure 3a plots the variation in temperatures on the day of the survey within these constructed climate zones. It demonstrates that there is sufficient temperature variation for the identification of temperature effects. Across the survey data collection period within climate zones individuals experience differences in temperature of several degrees. Finally, Figure 3b shows that the source of this temperature variation within climate zones is almost entirely due to the duration of time it took to roll-out the survey in each climate zone, rather than because of environmental or socioeconomic differences within the zones. We regress maximum temperature on the day of the survey on date of survey fixed-effects and find that 1.8% of climate zones have an $R^2$ between 0.66 and 0.8, 4.9% have an $R^2$ between 0.8 and 0.9, 14.9% have an $R^2$ between 0.9 and 0.95, and 78.4% have an $R^2$ above 0.95. These different $R^2$ values are represented in Figure 3b by colour (darker colours representing higher $R^2$ values). In a robustness test discussed in Section 4 we verify that our main results are nearly identical when using a subsample of climate zones with high $R^2$ values (i.e. zones in which practically all temperature variation is caused by variation in interview dates).

---

[13] The choice of thresholds for the altitude and distance to the coast groups was based on the shape of the distribution of both variables. The placement of the cut-off points is chosen so as to create approximately equally sized bins while maintaining a large enough number of observations within each threshold. Nevertheless, our main results are not sensitive to various adjustments of these thresholds.

[14] Interacting smaller administrative units than provinces, such as districts or subdistricts, significantly reduces the temperature variation available for identification (especially when considering lagged effects, as we do). Following standard procedures, the survey was rolled-out by geographical areas, and therefore some areas were completely surveyed within days. For instance, the sub-district of Bengo in the province of South Sulawesi was 96% surveyed in 12 consecutive days in 2015.

*3.2 Main Specification*

The effects of temperature on economic preferences are estimated with the following regression specification:

$$Pref_{iat} = \beta_1 temp_{at} + \beta_2 weather_{at} + \beta_3 lat_{ia} + \beta_4 X'_{iat} + \tau_t + \phi_a + \varepsilon_{iat} \qquad (1)$$

where $Pref_{iat}$ is the economic preference of individual *i*, residing in climate zone *a* in time *t*. Parameter $weather_{at}$ contains environmental variables specific to each village located in area *a*, at time *t* such as daily maximum precipitation, wind speed, and particulate matter 2.5 (pollution). We also include controls for village latitude which can cause temperatures to drop the further away from the equator. It has also been found to affect economic output (Hall & Jones, 1999), settlement patterns, and even disease environment (Mellinger et al., 2000), all of which could be linked to economic preferences, and therefore important to account for.

$X_{iat}$ consists of individual and household-level time-varying covariates such as age, gender, marital status, religion, the individual's main income-generating activity, their highest level of education, a squared function of equivalized expenditure, the same demographic characteristics for the household head, number of children, and numbers of adults in the house, day of the week, and hour when the interview began. Finally, $\tau_t$ and $\phi_a$ are a set of month-year and climate zone (province-urban-altitude-distance to coast) fixed effects, respectively. Month-year fixed effects control for potential seasonal patterns and festivities that could affect exposure to heat, cognition, sleep and economic preferences.[15]

---

[15] We cluster standard errors by village, the level of variation in our treatment variable (Abadie et al., 2017). However, our standard errors are similar if we change the level of clustering to a) the climate zone level, and b) the MERRA-2 grid level to account for smoothed averages at the grid level in weather station scarce cells

Our treatment variable *temp* represents maximum temperature on the day of the interview, midnight temperature the night before the interview, and the temperature at the hour when the survey began. We also consider the possibility that any potential effects could be non-linear. In a more flexible non-parametric specification, we employ bins of midnight temperature to study the shape of this relationship (<21ºC, 21-22, 22-23, 23-24, 24-25 and 26ºC >), with temperatures 14-21ºC as the reference. This reference level has been used in other studies and recognized as an ideal range for human comfort (Barreca et al., 2016; Connolly, 2013). There were no temperatures below 14ºC registered in the IFLS villages during the period of study, precluding a further split to study the effects of colder temperatures.

Analysis presented in the preceding section suggests that within climate zone variation in temperature is determined by interview date. However, we have also explored whether there is endogenous selection of village climate based on economic preferences. For the sample of people who moved villages between IFLS4 and IFLS5, we regress changes in a person's temperature between 2007-2008 and 2014-2015 on economic preferences measured in 2007-2008, and our standard set of covariates. The estimates in Appendix Table B2 show that economic preferences in 2007-2008 do not significantly predict maximum, mean and midnight temperatures in 2014-2015; supporting our identification assumptions.

## 4. Temperature Effects on Time and Risk Preferences

*4.1 Main Results*

In this section, we present the results from estimating regression equation (1). Table 3 shows estimates of the linear effects of maximum temperatures on the day of the survey (Panel A), the temperature when the survey began (Panel B), and midnight temperature the night before the survey (Panel C). The regressions in Panel D explore the relative importance of temperature on the day of the survey versus the night before.

We find that temperature, regardless of when it is measured, is unrelated to reported risk aversion (column 1, Table 3). This is consistent with the results in Almås, et al. (2020) obtained in laboratory-controlled environments. In contrast, we find that temperature has a significant positive relationship with rational choice violations and impatience. A 1ºC increase in the maximum temperature of the day of the survey is estimated to increase rational choice violations and impatience by 0.3 and 0.4 percentage points, respectively. Corresponding effects for temperature at the commencement of the survey equal 0.7 and 0.4 percentage points. The largest point estimates correspond to temperature on the midnight before the survey, with a 1ºC increase in temperature increasing rational choice violations by 1 percentage point and impatience by 0.9 percentage points. These latter effects are approximately equivalent to 30% and 70% of the gender gaps in these outcomes (see Appendix Table B1). The results in Panel D provide further evidence that high night-time temperatures are more impactful on economic preferences than high day-time temperatures.[16] When day-time and night-time temperatures are simultaneously included, the day-time temperature coefficients reduce in magnitude (and become statistically insignificant), while the midnight temperatures are very similar to the values in Panel C.

To test for non-linearity, we replace the continuous midnight temperature measure with five temperature categories. The estimated effects of these categories, relative to <21ºC, on the three preference outcomes are shown in Figure 4. The estimates for rational choice violations and impatience appear roughly linear, providing support for our main regression specification. When midnight temperatures are >26ºC it is estimated that respondents are 6.8 (22%) and 6.1 (10%) percentage points more likely to make rational choice violations and impatient decisions, respectively, compared to when midnight temperatures are <21ºC.

In Table 4, we test the sensitivity of the estimates to additional covariates. Columns 1 and 4 present results from regressions that include humidity controls (from the MERRA-2 dataset). Temperature and humidity are negatively correlated, and humidity can affect people's comfort levels. Columns 2 and 5 present results from regressions that include controls for longitude, a geographic coordinate that specifies the east-west position in reference to the prime meridian. Longitude does not directly affect climate, but it is the basis of time zones within Indonesia, which affects hourly weather records. Finally, Columns 3 and 6 present results from regressions with interviewer fixed effects. This accounts for the possibility that interviewers influence the timing of the survey and decisions made in the lottery games, and that interviewers themselves could be affected by heat. Our results remain relatively stable across the three alternative specifications and the *p-values* in Panel B show that the effects of maximum and midnight temperature remain significantly different from each other in most specifications.[17]

---

[16] To avoid multicollinearity issues in Panel D, we include only maximum temperature on the day of the survey and midnight temperature, which had the lowest pairwise correlation at approximately 0.35. In Appendix Table B4 we demonstrate that even when all three temperature variables are included together, midnight temperature still appears to be the most important for economic preferences.

*4.2 Effect Heterogeneity*

Next, we explore whether the observed effects of temperature on rational choice violations and impatience vary systematically by individual characteristics such as gender, age, and socioeconomic status, as well as household level expenditure in electricity in the month prior to the survey. Figure 5 shows that temperature effects on rational choice violations and impatience are not statistically different by gender or age group, but that there are significant differences by educational attainment (*p-values* = 0.004 and 0.632), consumption (*p-values* = 0.063 and 0.033), and electricity expenditure (*p-values* = 0.001 and 0.160).[18] For each significant difference, people with lower economic status (low education, low consumption and low electricity expenditure) are those with larger temperature effects. This result suggests that people of lower socio-economic status have less means to protect themselves from the effects of high temperatures and are, as a consequence, experiencing larger deteriorations in the quality of their decision-making abilities.

*4.3 Cumulative Effects*

In Table 5, we explore whether repeated nights of high temperatures could be having a cumulative effect leading to a larger change in economic preferences than a single hot night. We test this by regressing our main outcomes of interest on the number of nights above 25ºC in the past 7 days. The choice of this threshold is based on the results in Figure 4 that show significant positive effects above this value. This choice is also supported by the findings of Minor et al. (2020) that night-time minimum temperatures above 25ºC significantly increase the probability of getting less than 7 hours of sleep. We also interact a dummy for temperatures above 25ºC on the night prior to the survey and the number of hot nights in the past seven days. This allows last night's temperature to have a different effect depending upon the number of nights in the past week that were also hot.

---

[17] A full set of sensitivity checks including results using maximum temperature on the day of the survey and temperature at the start of the survey as treatment variables can be found in Appendix Table B3.
[18] Household electricity expenditure is used as a proxy for air-conditioner use at home. Households with low energy consumption are unlikely to be using an air-conditioner.

The estimates in column 1 of Table 5 suggest that having a hot night directly before the survey and the number of hot nights during the week prior both increase rational choice violations. We observe a similar effect for impatience (column 2). The estimated coefficient for a hot night yesterday becomes statistically insignificant, but each hot night in the past week significantly increases impatience by 0.5 percentage points. The coefficient on the interaction term is negative in both regressions, indicating that the positive effect of a hot night before the survey is smaller if it follows a hot week; however, these coefficients are imprecisely estimated.

*4.4 Robustness Checks*

We estimate several additional regressions to determine whether our main results are robust to alternative specifications and sample exclusions. First, we include covariates indicating whether individuals were surveyed during Ramadan. During this period, those who observe the custom normally change their eating and sleeping patterns, and these changes could impact economic preferences. Only 1.6% of our sample was surveyed during Ramadan, and so this is unlikely to be a major driver of our results. Nevertheless, we test for this possibility and the results show that the additional controls have no effect on the magnitude or significance of our original estimates (see Appendix Table B4). The coefficient on the Ramadan term is close to zero and negative for rational choice violations but large and positive for impatience. However, it is statistically insignificant in both cases. This indicates that either the period of time when the custom is observed has no incidence on the occurrence of rational choice violations and impatience, or that our main specification already accounts for its effects.

Second, we exclude survey respondents who were visited more than once to complete the survey. Multiple visits could be (partly) caused by temperature – for instance, if interviewees suspend the interview until a cooler, more comfortable time – and incidence of multiple visits may be associated with individual characteristics associated with economic preferences. Results in Appendix Table B4 indicate that our main estimates are robust to the exclusion of the 19% of the sample with more than one visit.

Third, we exclude all observations in climate zones where less than 90% of the variation in temperature is explained by date of survey fixed-effects (see discussion in Section 3.1 and associated Figure 3B). In these climate zones, part of the variation in temperature may be caused by differences in environmental and geographical factors, which may not be exogenous in our regressions. In Appendix Table B5, we demonstrate that excluding these climate zones has very little effect on the estimated coefficients.

Next, we perform a placebo test using future temperatures. In Appendix Table B6, we add the maximum temperature of the same day-of-the-week, but two weeks later. Being in the future, this variable should have no statistically or economically significant effect on any of our preference outcomes, so long as the current temperature and the temperature in two weeks is not highly correlated. Results indicate that the estimates for midnight temperatures are unchanged and the effects of future temperature on rational choice violations and impatience are close to zero and statistically insignificant. Finally, we perform a second placebo test by regressing individual and household characteristics that should not be affected by weather on our main measures of temperature. Appendix Table 7 shows, as expected, that midnight temperature on the night prior to the survey and temperature on the day of the survey do not significantly affect the respondent's age, gender, marital status or even household equivalised expenditure in the past month.

## 5. Potential Mechanisms

The relationship between night-time temperature and economic preferences may be driven by many different mechanisms. Three that are especially likely are cognition, sleep, and mood. In the following sub-sections we empirically explore the effects of temperature on these three outcomes, and evaluate the likelihood that they are important mediating factors.

*5.1 Temperature Effects on Cognition*

Several studies have studied the links between cognition and economic preferences with mixed results. For instance, Benjamin et al. (2013) find that students with high math test scores demonstrate low risk aversion in small-stakes gambles and less short-run discounting. Similarly, Falk et al. (2018) find that math skills are "uniformly positively linked to patience, risk taking, and social preferences" (p. 1647). In contrast, Andersson et al. (2016) find that cognitive ability is related to errors in decision making but not to risk preferences. At the same time, a growing body of evidence shows that high temperatures have the potential to impair cognition (Zivin et al., 2018; Garg et al., 2020; Park et al., 2020a; 2020b). Together, this evidence suggests that high temperatures deplete cognitive function, which then in turn, affects economic preferences.

In Panel A of Table 6 we study the effects of night-time temperature on cognitive functioning. Results are presented separately for waves 4 and 5 because of the differences in the ages of respondents tested (15-31 in wave 4; 15-90 in wave 5). The top row contains results for the predicted factor from a principal component analysis of all available cognitive tests. The second row contains results for the standardized score from the math test, which is included because past studies have emphasized the importance of math skills for economic preferences. The estimates show that a 1°C increase in midnight temperature on the night before the tests decreases overall cognition by 3.6% of a standard deviation and 1.3% of a standard deviation for individuals aged 15-31 in IFLS4 and individuals aged 15-90 in IFLS5, respectively.[19] Results are similar for the standardized math score: a 1°C increase reduces scores by 3.9% and 1.3% of a standard deviation.[20]

In light of these results, we have also explored the association between cognitive functioning and economic preferences. Interestingly, we find that a one standard deviation increase in overall cognition is not associated with risk aversion (0.7 percentage points), but is strongly associated with reduced rational choice violations (5.3 percentage points) and impatience (5.5 percentage points); conditional on age, sex, educational attainment, household demographics, and household expenditure (see Appendix Table B1). Though these estimates cannot be interpreted causally, the fact that temperature affects cognition, rational choice violations and impatience (but not risk aversion), and that cognition is associated with rational choice violations and impatience (but not risk aversion), is highly suggestive of cognition being an important pathway for the temperature effects shown in Table 3. Consistent with the results in Figure 4, Appendix Figure A4 shows that there is a nearly monotonic (decreasing) relationship between night-time temperatures and cognition in both IFLS waves.

*5.2 Temperature Effects on Sleep*

Sleep is another potential mechanism for the relationship between temperature and preferences. Better sleep quality is associated with lower discount rates, less present bias, and less loss aversion (Nofsinger & Shank, 2019). Conversely, lack of sleep has been shown to increase financial risk taking (McKenna, et al., 2007; Killgore, 2007; Killgore et al., 2012). Sleep has also been shown to be affected by higher temperatures. For example, Minor et al. (2020) show that rising night-time temperatures shorten sleep duration by delaying its onset, and increase the probability of insufficient sleep. Similarly, Obradovich et al. (2017) show that higher nighttime temperatures increase self-reported nights of insufficient sleep, such that a 1°C night-time temperature anomaly produces an increase of 3 nights of insufficient sleep per 100 individuals per month.[21]

---

[19] In Appendix Table B8, we re-estimate the cognition effects for respondents aged 15-31 in IFLS5 so that we can compare effect sizes across waves. The estimates indicate much larger (more negative) temperature effects in wave 4 than in wave 5. These findings could suggest that our sample is increasingly adapting to high temperatures, for example by acquiring air-conditioning units.
[20] For wave 4, temperature also significantly decreased scores on the date awareness and word recall test (3.3%). For wave 5, temperature also significantly decreased the fluid intelligence score (1.6%) and the Raven's Progressive Matrices test score (2.0%).

Panel B of Table 6 reports the estimated effects of midnight temperature on sleep onset (the time when individuals went to bed), sleep offset (the time when individuals woke up), and the subtraction of these two, which produces a measure of how many hours an individual spent in bed. Each estimate is small and statistically insignificant. For instance, the point estimate for sleep duration implies a 0.42-minute increase for a 1ºC increase in temperature. This suggests that sleep length is not an important mediating factor. An important caveat, however, is that our crude self-reported measures of sleep, fail to capture the percentage of time asleep in relation to time spent in bed, known as sleep efficiency (Lauderdale et al., 2008). More accurate measures of sleep quantity and quality, such as those derived from accelerometer-based wearable devices and detailed validated sleep questionnaires, should be used to corroborate this finding.

*5.3 Temperature Effects on Mood*

The final potential mechanism we investigate is mood. The empirical evidence shows that temperature can significantly influence our mood. For example, a number of studies have documented that exposure to high temperatures can cause extreme responses such as violent acts of aggression in multiple settings and agents, from American football and baseball players (Craig et al., 2016; Krenzer & Splan, 2018), to prison inmates (Mukherjee & Sanders, 2021), and university students (Almås et al., 2020). Others have studied different emotions such as revealed happiness and found that individuals living through a very hot day can experience a decrease in happiness similar to a happiness drop from a Sunday to a Monday (Baylis, 2020). These findings are relevant to our study because mood can significantly influence economic preferences. For example, aggression is positively associated with risk-taking behavior (Cueva et al., 2015; Deffenbacher et al., 2003; Arnett et al., 1997), while a happy mood is associated with significant reductions in impatience over money (Ifcher & Zarghamee, 2011). The relationship between happiness and risk preferences is mixed with some studies finding that positive mood leads to risk seeking behavior (Campos-Vasquez & Cuilty, 2014), and others finding that positive mood leads to more risk-averse behavior (Kliger & Levy, 2003). However, support has more often been found for the former (Lane, 2017).

---

[21] In Obradovich et al. (2017) an anomaly is the 30-day average of daily minimum temperature deviations from their normal daily values (1981-2010) over the same 30-day period when respondents report insufficient sleep.

We explore the effects of temperature on mood in panel C of Table 6 and find that midnight temperatures prior to the survey have no significant effect on the likelihood of feeling anger, tiredness, happiness, or enthusiasm. These findings suggest that reduced cognition and not sleep or mood, is the most likely channel through which high night-time temperatures are affecting rational choices and impatience.

## 6. Discussion

Using Indonesian data eliciting economic preferences in combination with climate records from NASA's MERRA-2 dataset, we estimate the causal impacts of temperature on time and risk preferences, and rational choice. Identification is based on comparing survey responses between people living in the same climate zone, who experienced different temperatures because of the allocation of different survey dates. Through this approach we add to the small literature on the causal effects of temperatures on economic preferences, which mostly contains evidence from laboratory experiments.

Results show that higher temperatures on the day of the survey and the night prior to the survey significantly increase the likelihood of making rational choice violations and making impatient choices. Night-time temperatures are especially important, with a 1°C increase in midnight temperature associated with a 3.2% (relative to the sample mean) increase in irrational behavior and a 1.5% increase in impatience. Results also indicate a model of cumulative heat effects, such that multiple nights of high temperatures in the preceding week add to the effect from the night before the survey. Finally, reduced cognitive functioning seems the most likely driver of these results. High night-time temperatures significantly reduce cognitive functioning the next day, with mathematical skills particular affected, while self-reported sleep duration and mood are unaffected.

Altogether, these findings indicate that heat-induced night-time disturbances appear to cause stress on the brain, leading to significantly lower cognitive functions the following day. These lower cognitive skills in turn, prevent individuals from engaging their system 2 thought processes to perform economically rational decision-making. As Cheema & Patrick (2012) suggest, it appears as though the physical and psychological stress that heat imposes upon individuals, leads them to rely more often on intuition and lower-level processing, also known as system 1 thinking. This, in turn, is reflected in a substantial increase in sub-optimal economic decision making as individuals default towards the certain option and the earlier albeit smaller reward.

Importantly, a decomposition of temperature effects by individual characteristics shows that the effects are larger for poorer households, likely due to their lower ability to adapt, such as by acquiring air-conditioning units. This indicates that the effects of a rise in the intensity and frequency of extreme heat will be spread unevenly. Effects will likely differ between low- and high-income households within countries, and between developed and developing countries. Greater macroeconomic and household wealth mean that air-conditioning coverage across the developed world is becoming widespread in homes, office spaces, and public transport, such that individuals in wealthier nations can mostly protect themselves from the deleterious impacts of heat. This is evidenced in the 75% decline in heat-induced fatalities in the United States since the 1960's; almost entirely due to residential air-conditioning (Barreca et al., 2016).

Across the developing world, the financial and infrastructural capacity to deal with high temperatures is significantly lower. Dwellings need to be connected to a power grid, yet 13% of the world does not have access to electricity (World Bank, 2021), and households need to have the income to acquire air-conditioning units and pay the associated electricity bills. Furthermore, low-cost dwellings tend to be built with materials that store heat rather than insulate their inhabitants from it (Naicker et al., 2017). Overall, further attention and research is needed on how the poor make investment and risky decisions, and whether or not these are systematically less optimal in regions of the world where temperatures are getting closer to the upper threshold of human adaptability.

Cameron, L., & Shah, M. (2015). Risk-taking behavior in the wake of natural disasters. *Journal of Human Resources, 50*(2), 484-515. doi:10.3368/jhr.50.2.484

Campos-Vasquez, R. M., & Cuilty, E. (2014). The role of emotions on risk aversion: A Prospect Theory experiment. *Journal of Behavioral and Experimental Economics, 50*, 1-9. doi:https://doi.org/10.1016/j.socec.2014.01.001

Carvalho, L. S., Meier, S., & Wang, S. W. (2016). Poverty and Economic Decision-Making: Evidence from Changes in Financial Resources at Payday. *American Economic Review, 106*(2), 260-284. doi:http://dx.doi.org/10.1257/aer.20140481

Cheema, A., & Patrick, V. M. (2012). Influence of Warm Versus Cool Temperatures on Consumer Choice: A Resource Depletion Account. *Journal of Marketing Research, 49*(6), 984-995. doi:https://doi.org/10.1509/jmr.08.0205

Conlin, M., O'Donoghue, T., & Vogelsang, T. J. (2007). Projection bias in catalog orders. *American Economic Review, 97*(4), 1217-1249. doi:10.1257/aer.97.4.1217

Connolly, M. (2013). Some like it mild and not too wet: The influence of weather on subjective well-being. *Journal of Happiness Studies, 14*(2), 457-476. doi:https://doi.org/10.1007/s10902-012-9338-2

Craig, C., Overbeek, R. W., Condon, M. V., & Rinaldo, S. B. (2016, June). A relationship between temperature and aggression in NFL football penalties. *Journal of Sport and Health Science, 5*(2), 205-210. doi:https://doi.org/10.1016/j.jshs.2015.01.001

Cueva, C., Roberts, R. E., Spencer, T., Rani, N., Tempest, M., Tobler, N. P., . . . Rustichini, A. (2015). Cortisol and testosterone increase financial risk taking and may destabilize markets. *Nature Scientific Reports, 5*(11206). doi:10.1038/srep11206

Deffenbacher, J. L., Deffenbacher, D. M., Lynch, R. S., & Richards, T. L. (2003). Anger, aggression, and risky behavior: a comparison of high and low anger drivers. *Behavior Research and Therapy, 41*, 701-718. doi:10.1016/S0005-7967(02)00046-3

Ekelund, J., Johansson, E., Marjo-Riitta, J., & Dirk, L. (2005). Self-employment and risk aversion—evidence from psychological test data. *Labour Economics, 12*, 649 – 659. doi:10.1016/j.labeco.2004.02.009

EPA. (2021). *Reduce Urban Heat Island Effect*. Retrieved August 08, 2021, from United States Environmental Protection Agency: https://www.epa.gov/green-infrastructure/reduce-urban-heat-island-effect

Falk, A., Becker, A., Dohmen, T., Enke, B., Huffman, D., & Sunde, U. (2018). Global Evidence on Economic Preferences. *Quarterly Journal of Economics, 133*(4), 1645-1692. doi:10.1093/qje/qjy013

Fong, T. G., Fearing, M. A., Jones, R. N., Shi, P., Marcantonio, E. R., Rudolph, J. L., . . . Inuoye, S. K. (2009). The Telephone Interview for Cognitive Status: Creating a crosswalk with the Mini-Mental State Exam. *Alzheimers Dement., 5*(6), 492-497. doi:10.1016/j.jalz.2009.02.007

Garg, T., Jagnani, M., & Taraz, V. (2020). Temperature and Human Capital in India. *Journal of the Association of Environmental and Resource Economists, 7*(6), 1113-1150. doi:https://doi.org/10.1086/710066
22

# Figures

## Figure 1. IFLS5 villages and MERRA-2 grids

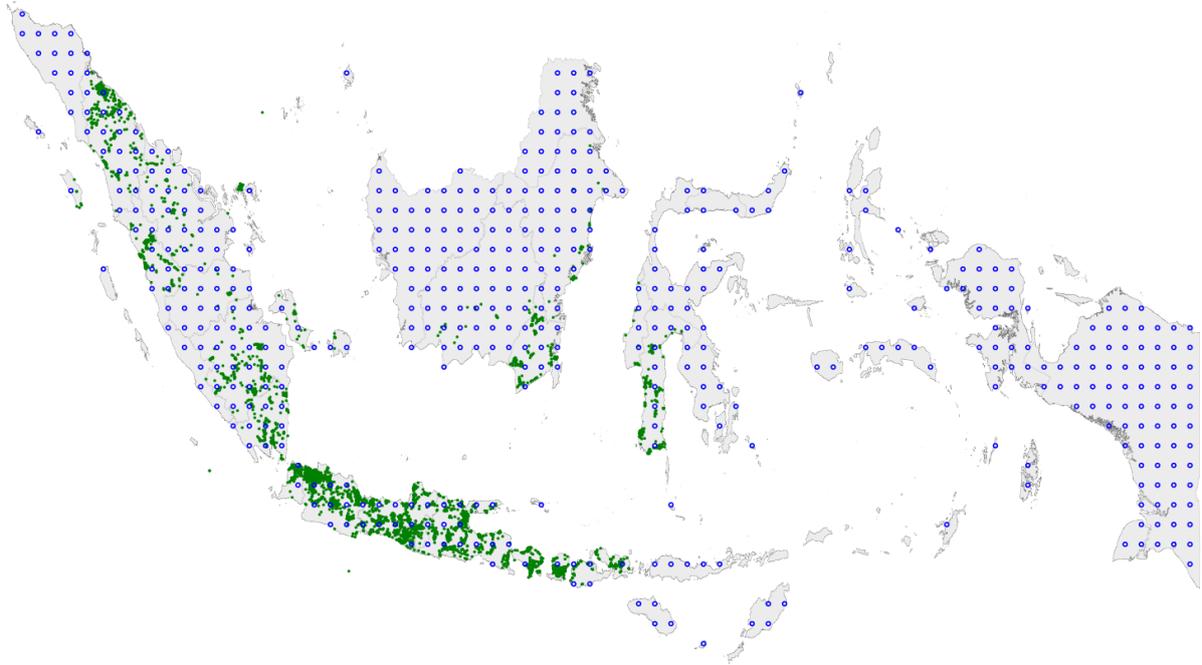

**Notes:** The above figure shows a map of the Indonesian territory plotting NASA's MERRA-2 grids represented by the blue circles as well as the villages covered in IFLS5 represented by the green dots. Map created by authors.

**Figure 2. Map of IFLS villages in North Sumatra by constructed climate zones**

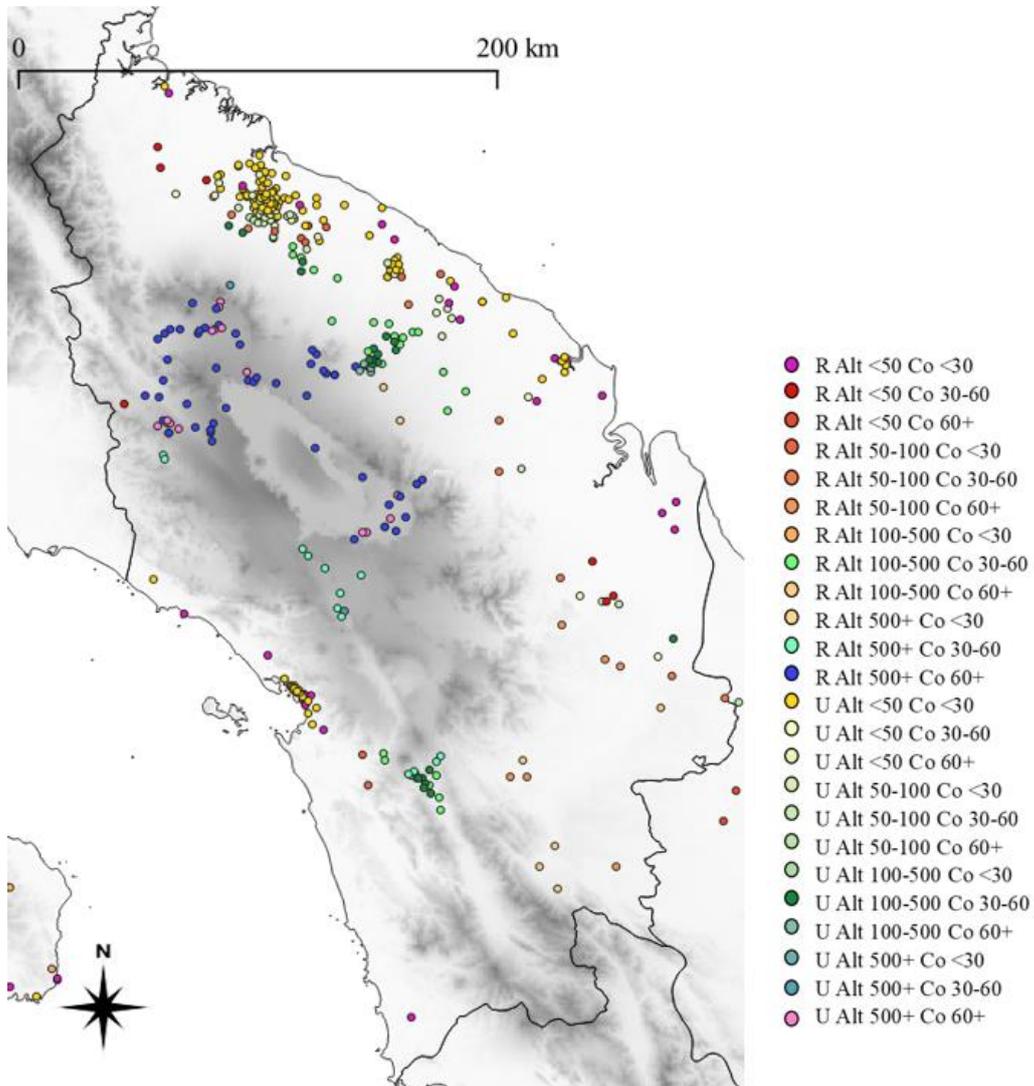

**Notes:** The above figure shows the map of the Indonesian province of North Sumatra with elevation in the background and IFLS villages in bins grouped by urban or rural, 4 altitude groups (meters) and 3 distance to the coast groups (kilometers) resulting in 19 sub-groups for this province.

**Figure 3. Within climate zone temperature variation**

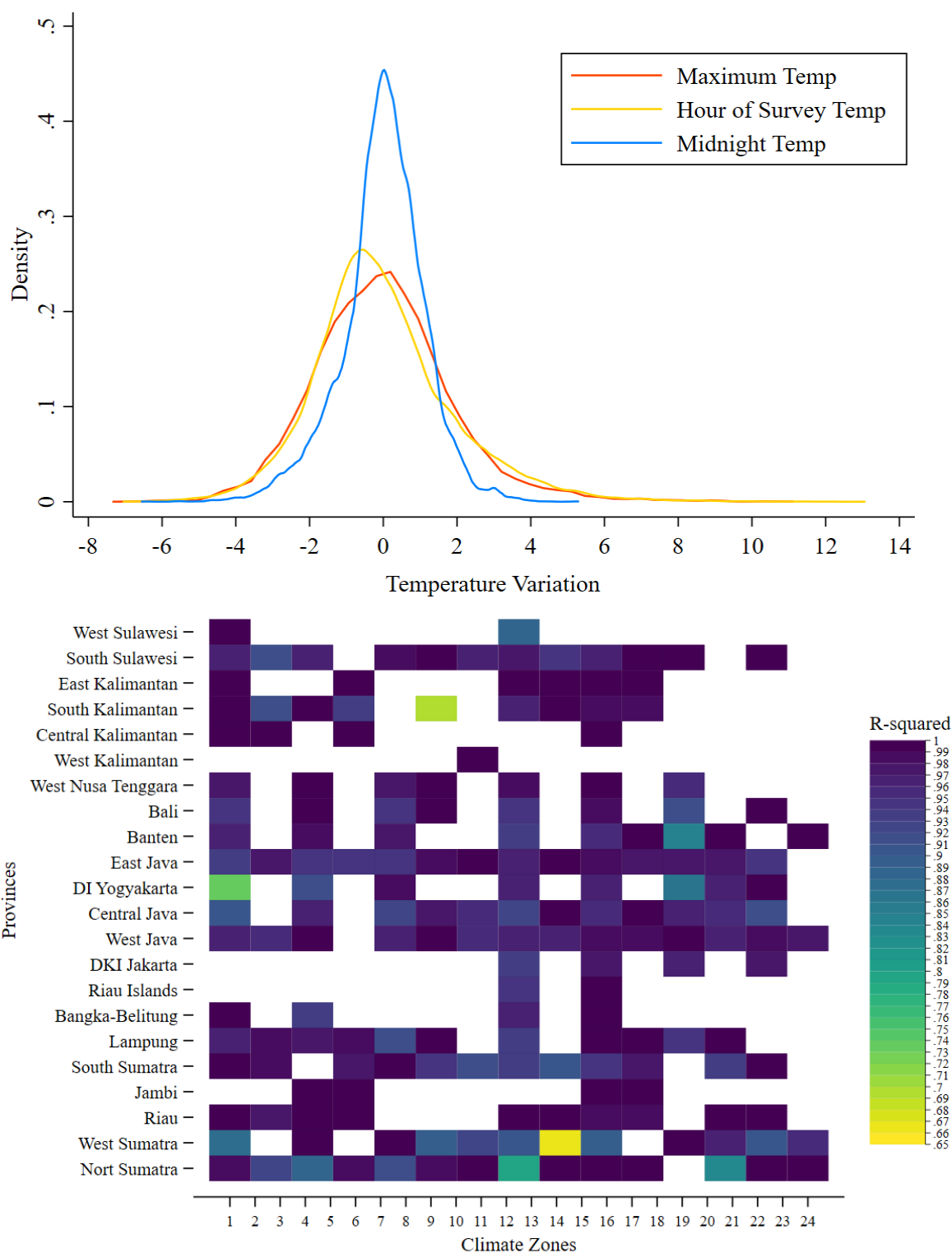

**Notes:** Figure 3a (top) shows the density functions of maximum temperature on the day of the survey, at the hour when the surveys began and on the midnight before the survey, using a bandwidth of 0.1688. The with-in bin variation is the result of subtracting from each individual's temperature, the average temperature in their corresponding bin. Figure 3b (bottom) shows the proportion of the variation in maximum temperature ($R^2$) explained by date-of-survey fixed effects in individual regressions for every climate zone. Blank squares are either climate zones that do not exist in the province, zones where no people were surveyed, or zones with too few observations to execute the regression.

**Figure 4. Nonlinear effects of *midnight temperatures* on time and risk preferences**

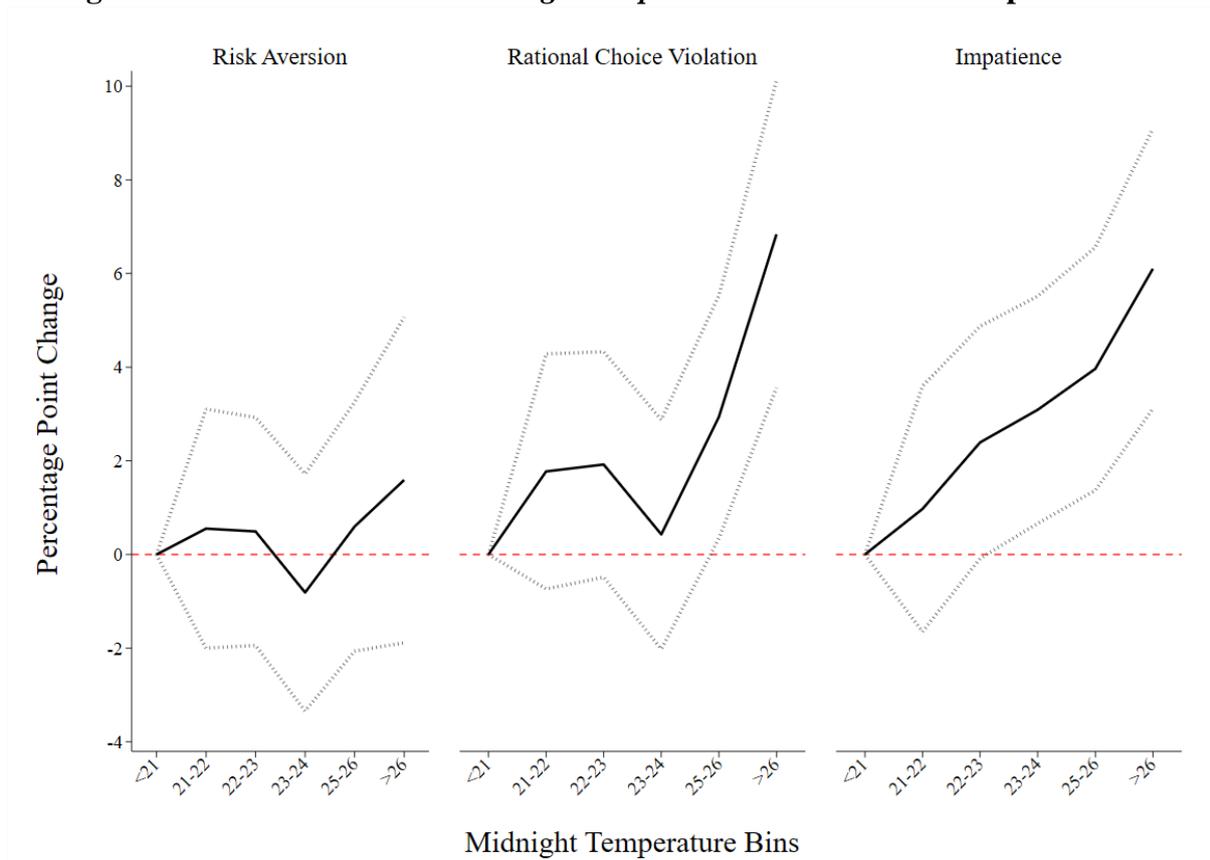

**Notes:** The above figure shows nonlinear effects of midnight temperature bins on the probability of making a rational choice violation, being in the most risk averse category, and in the most impatient category.

**Figure 5. Heterogeneous effects of *midnight temperatures* by sub-samples**

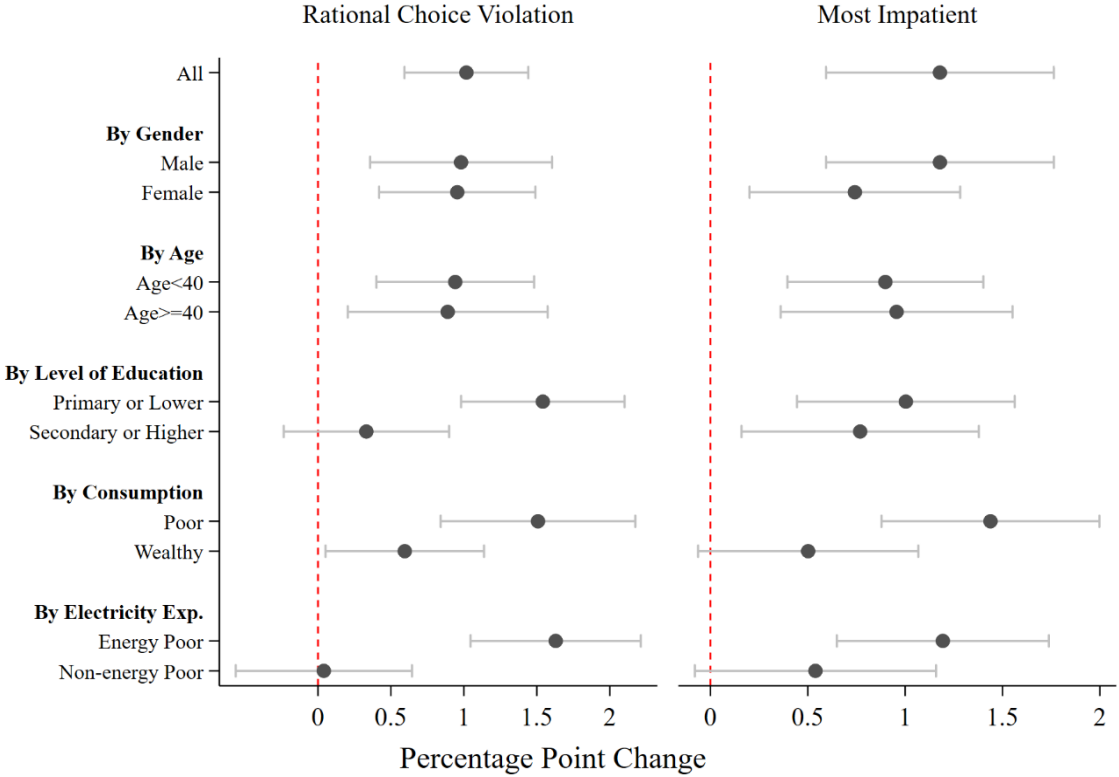

**Notes:** The above figure shows heterogeneous effects of temperature at midnight the day before the survey on rational choice violations and impatience in percentage points, by gender, age groups, level of education, and equivalized consumption expenditure.

# Tables

## Table 1. Key variables used in IFLS regression analyses

| Variables | Description | Mean | SD |
|---|---|---|---|
| **(A) Treatment** | | | |
| Max temperature on day of survey (ºC) | Maximum temperature on the day of the survey | 28.67 | 2.24 |
| Temperature at start of survey (ºC) | Average temperature at hour when the survey started | 24.55 | 2.35 |
| Midnight temperature yesterday (ºC) | Temperature at midnight the day before the survey | 23.64 | 1.85 |
| Midnight yesterday above 25ºC | Binary indicator if midnight prior to survey was 25ºC+ | 0.24 | 0.43 |
| # Nights above 25ºC | Number of midnights above 25ºC in the past 7 days | 1.65 | 2.66 |
| **(B) Standard Economic Preferences** | | | |
| Risk Aversion | Binary indicator for highest level of risk aversion excluding rational choice violations | 0.37 | 0.48 |
| Rational Choice Violation | Individual chooses a) IDR 800K over b) a coin toss between IDR 1.6 M or IDR 800 K | 0.31 | 0.46 |
| Impatience | Binary indicator for highest level of present bias from a ladder-type lottery game | 0.62 | 0.49 |
| **(C) Mechanisms** | | | |
| Cognition Z-Score | PCA score of Fluid Intelligence, TICS, Raven's Matrix and Math | 0 | 1 |
| Time in Bed | Total hours an individual spent in bed (sleep offset - sleep onset) | 6.8 | 1.88 |
| Angry | Binary indicator for a person who felt a little, somewhat, quite a bit or very angry yesterday | 0.31 | 0.46 |
| Tired | Binary indicator for a person who felt somewhat, quite a bit or very tired yesterday | 0.45 | 0.50 |
| Enthusiastic | Binary indicator for a person who felt quite a bit or very enthusiastic yesterday | 0.58 | 0.49 |
| Happy | Binary indicator who a person who felt quite a bit or very happy yesterday | 0.64 | 0.48 |

**Notes:** All treatment variables are created using data from NASA's Modern-Era Retrospective Analysis for Research and Applications, Version 2. Outcome variables for risk aversion, rational choice violations, and impatience use the combined IFLS4 + IFLS5 sample. Sleep, and mood outcomes use exclusively IFLS5 data. Cognition outcomes use both IFLS4 and IFLS5 data.

# Table 2. Accumulation decisions, risky choices and economic preferences

|  | Accumulation decisions | | Risky Choices | | | |
| --- | --- | --- | --- | --- | --- | --- |
|  | Savings | Higher Education | Self Employed | Plan to Start Business | Use Tobacco | Number Cigarettes |
|  | (1) | (2) | (3) | (4) | (5) | (6) |
| Impatience | -0.033*** | -0.088*** |  |  |  |  |
|  | (0.004) | (0.004) |  |  |  |  |
| Risk Aversion |  |  | -0.013*** | -0.010*** | -0.000 | -0.298** |
|  |  |  | (0.005) | (0.003) | (0.004) | (0.148) |
| Rational Choice Violation |  |  | -0.006 | -0.012*** | 0.003 | -0.301** |
|  |  |  | (0.004) | (0.003) | (0.004) | (0.140) |
| Outcome Mean | 0.21 | 0.48 | 0.27 | 0.07 | 0.34 | 11.65 |
| Observations | 38,463 | 52,432 | 51,778 | 49,190 | 49,574 | 16,227 |

**Notes:** All models are estimated using province-urban, month and year fixed effects. They also include controls for respondents' age, gender, marital status, religion, employment status, education (except for column 2) and household characteristics such as number of people and children in the house and equivalized monthly household expenditure. Robust standard errors are in parentheses. * $p<0.1$; ** $p<0.05$; *** $p<0.01$.

## Table 3. Linear effects of temperature on time and risk preferences

|  | Risk Aversion | Rational Choice Violation | Impatience |
|---|---|---|---|
|  | (1) | (2) | (3) |
| **Panel A** | | | |
| Max Temperature on Day of Survey (ºC) | -0.000 | 0.003 | 0.004** |
|  | (0.002) | (0.002) | (0.002) |
| **Panel B** | | | |
| Temperature during hour of survey (ºC) | 0.001 | 0.007*** | 0.004** |
|  | (0.002) | (0.002) | (0.002) |
| **Panel C** | | | |
| Midnight Temperature Yesterday (ºC) | 0.000 | 0.010*** | 0.009*** |
|  | (0.003) | (0.003) | (0.003) |
| **Panel D** | | | |
| Max Temperature on Day of Survey (ºC) | -0.000 | 0.001 | 0.002 |
|  | (0.002) | (0.002) | (0.002) |
| Midnight Temperature Yesterday (ºC) | 0.001 | 0.009*** | 0.009*** |
|  | (0.003) | (0.003) | (0.003) |
| p-value | 0.7904 | 0.0480 | 0.0763 |
|  | | | |
| Outcome Mean | 0.37 | 0.31 | 0.62 |
| Observations | 32,119 | 48,953 | 47,566 |

**Notes:** The p-values in panel D, test whether the coefficients for maximum temperature on the day of the survey and midnight temperature are significantly different from each other. All models include climate zone fixed effects which are the result of interacting province of residence, urbanity of the village, altitude groups, and distance to coast groups, as well as month and year fixed effects, and village latitude. The results exclude individuals living more than 50 kms from the closest grid. All models control for individual and household characteristics. Individual controls include: gender, age, marital status, income generating activity, marital status, religion, highest level of education, day of the week when the individual's survey took place and hour when the survey started. Household controls include: gender of the household head, their age, marital status, education level, number of children and adults in the household, and a squared function of equivalized consumption. We also include linear controls for pollution (PM 2.5), precipitation and wind speed. Columns 1-3 are estimated on the combined IFLS4 + IFLS5 sample. Robust standard errors clustered on village level are in parentheses. * p<0.1; ** p<0.05; *** p<0.01.

**Table 4. Alternative specifications - Effects of temperature on time and risk preferences**

|  | Rational Choice Violation | | | Impatience | | |
|---|---|---|---|---|---|---|
|  | (1) | (2) | (3) | (4) | (5) | (6) |
| **Panel A** | | | | | | |
| Midnight Temperature Yesterday (ºC) | 0.010*** | 0.008*** | 0.007*** | 0.009*** | 0.009*** | 0.006*** |
|  | (0.002) | (0.002) | (0.002) | (0.002) | (0.002) | (0.002) |
| **Panel B** | | | | | | |
| Max Temperature on Day of Survey (ºC) | 0.001 | 0.001 | 0.000 | 0.002 | 0.002 | -0.001 |
|  | (0.002) | (0.002) | (0.002) | (0.002) | (0.002) | (0.002) |
| Midnight Temperature Yesterday (ºC) | 0.009*** | 0.008*** | 0.007*** | 0.009*** | 0.008*** | 0.007*** |
|  | (0.002) | (0.002) | (0.002) | (0.002) | (0.002) | (0.002) |
| p-value | 0.0180 | 0.0898 | 0.0417 | 0.3335 | 0.1165 | 0.0459 |
|  | | | | | | |
| Outcome Mean | 0.31 | 0.31 | 0.31 | 0.62 | 0.62 | 0.62 |
| Observations | 48,953 | 48,953 | 48,953 | 47,566 | 47,566 | 47,566 |
|  | | | | | | |
| Humidity | Y | | | Y | | |
| Longitude | | Y | | | Y | |
| Interviewer FE | | | Y | | | Y |

**Notes:** The p-values in panel D, test whether the coefficients for maximum temperature on the day of the survey and midnight temperature are significantly different from each other. All models include climate zone fixed effects which are the result of interacting province of residence, urbanity of the village, altitude groups, and distance to coast groups. The results exclude individuals living more than 50 kms from the closest grid. All models control for individual and household characteristics and also include linear controls for pollution, precipitation and wind speed. Individual controls include: gender, age, marital status, income generating activity, marital status, religion, highest level of education, day of the week when the individual's survey took place and hour when the survey started. Household controls include: gender of the household head, their age, marital status, education level, number of children and adults in the household, and a squared function of equivalized consumption. Columns 1-6 are estimated on the combined IFLS4 + IFLS5 sample. Robust standard errors clustered on village level are in parentheses. * $p<0.1$; ** $p<0.05$; *** $p<0.01$.

## Table 5. Cumulative effects of midnight temperature

|  | Rational Choice Violation | Impatience |
|---|---|---|
|  | (1) | (2) |
| **Panel A - Contemporaneous Effects** |  |  |
| Midnight Temp Yesterday above 25ºC | 0.023** | 0.013* |
|  | (0.009) | (0.008) |
| **Panel B - Cumulative Effects** |  |  |
| Midnight Temp Yesterday above 25ºC | 0.022* | 0.016 |
|  | (0.012) | (0.012) |
| # Midnights above 25ºC Last Week | 0.007** | 0.005* |
|  | (0.003) | (0.003) |
| Interaction of above temperature variables | -0.007 | -0.006 |
|  | (0.005) | (0.004) |
| F-test | 2.87 | 1.80 |
| p-value | 0.0567 | 0.1655 |
|  |  |  |
| Outcome Mean | 0.31 | 0.62 |
| Observations | 48,953 | 47,566 |

**Notes:** Columns 1 and 2 include data from both IFLS4 and IFLS5. All models include climate zone fixed effects, the result of interacting province of residence, urbanity of the village, altitude groups and distance to coast groups, plus month and year fixed effects and linear controls for pollution, precipitation and wind speed. All regressions exclude individuals living more than 50 kms from the closest grid. We include the p-value of the test of joint significance for the cumulative temperature variables: # of midnights with temperature above 25ºC and the interaction term in Panel B. Robust standard errors clustered on village level are in parentheses.
 * $p<0.1$; ** $p<0.05$; *** $p<0.01$.

**Table 6. Potential mechanisms behind temperature effects**

|  | Wave 4 |  | Wave 5 |  |
|---|---|---|---|---|
| **Panel A** – Cognition |  |  |  |  |
| Cognition factor z-score | -0.036** | (0.017) | -0.016** | (0.006) |
| Math z-score | -0.039*** | (0.014) | -0.013** | (0.006) |
|  |  |  |  |  |
| **Panel B** – Sleep |  |  |  |  |
| Sleep Onset (hours) | - |  | -0.003 | (0.011) |
| Sleep Offset (hours) | - |  | 0.004 | (0.011) |
| Sleep Duration (hours) | - |  | 0.007 | (0.011) |
|  |  |  |  |  |
| **Panel C** – Mood |  |  |  |  |
| Angry (0/1) | - |  | -0.002 | (0.003) |
| Tired (0/1) | - |  | -0.004 | (0.003) |
| Enthusiastic (0/1) | - |  | -0.003 | (0.003) |
| Happy (0/1) | - |  | -0.003 | (0.003) |

**Notes:** Outcomes in panels A and B are regressed on temperature the night before the survey. Panel C outcomes use 'yesterday' as reference so they are regressed on midnight temperature the day before yesterday. Cognition measures in IFLS4 were only collected for individuals age 15-31. IFLS5 cognition measures include the full sample of individuals age 15-90. Sleep and mood variables were not collected in IFLS4. All regressions include climate zone fixed effects, month and year fixed effects, controls for latitude, pollution (PM2.5), rain and wind speed.
Robust standard errors clustered on village level are in parentheses.
 * $p<0.1$; ** $p<0.05$; *** $p<0.01$.

# Appendix A – Figures

## Appendix Figure A1. Risk preferences flowchart

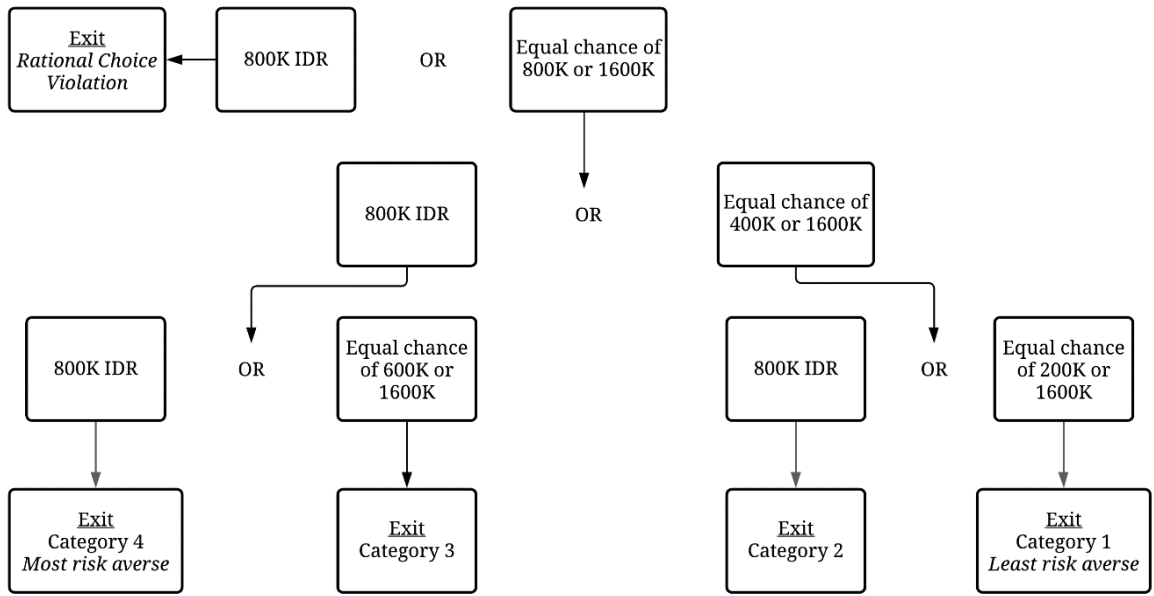

**Notes:** The above figure shows a flowchart adapted from IFLS4 and IFLS5 illustrating the elicitation of risk preference module based on figure by Ng (2013).

## Appendix Figure A2. Time preferences flowchart

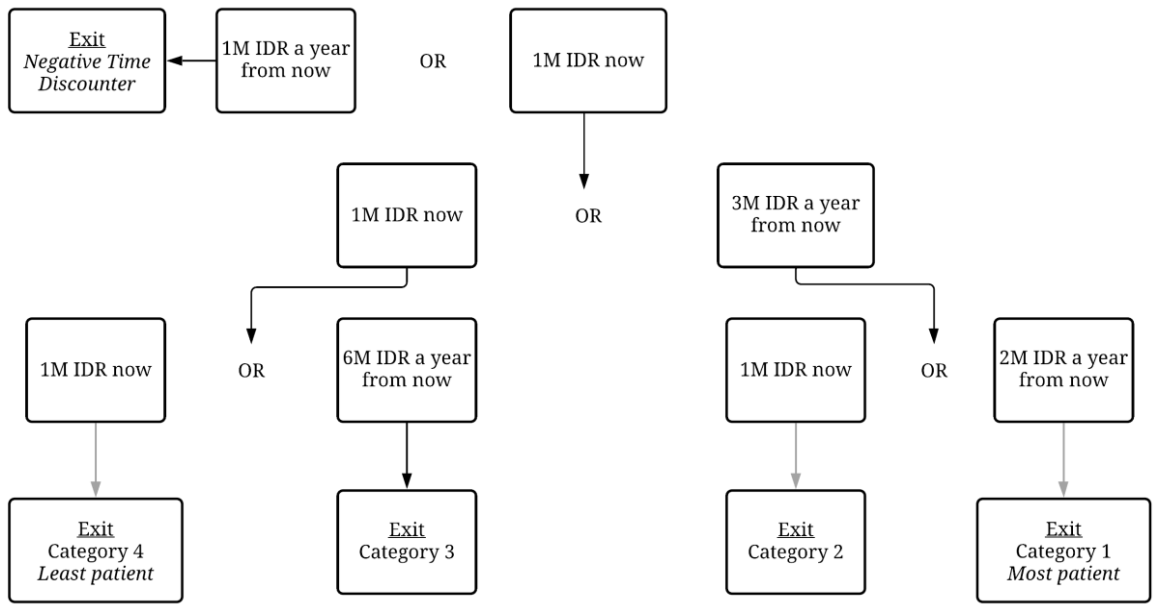

**Notes:** The above figure shows a flowchart adapted from IFLS4 and IFLS5 illustrating the elicitation of time preference module based on figure by Ng (2013).

**Appendix Figure A3. Number of days to roll out survey by climate zone**

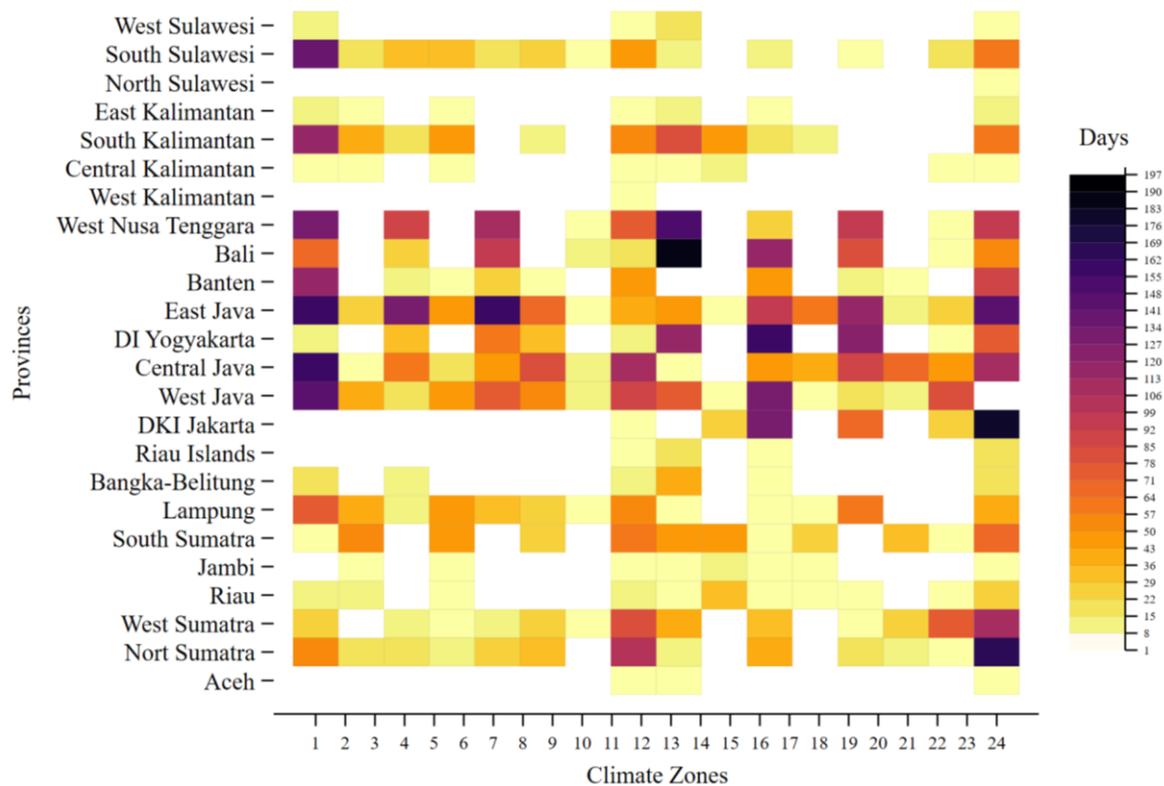

**Notes:** The figure above shows the total number of days it took to rollout the IFLS survey in each climate zone. Blank squares are either climate zones that do not exist in the province or zones where no people were surveyed.

**Appendix Figure A4. Effects of *midnight temperatures* on cognition**

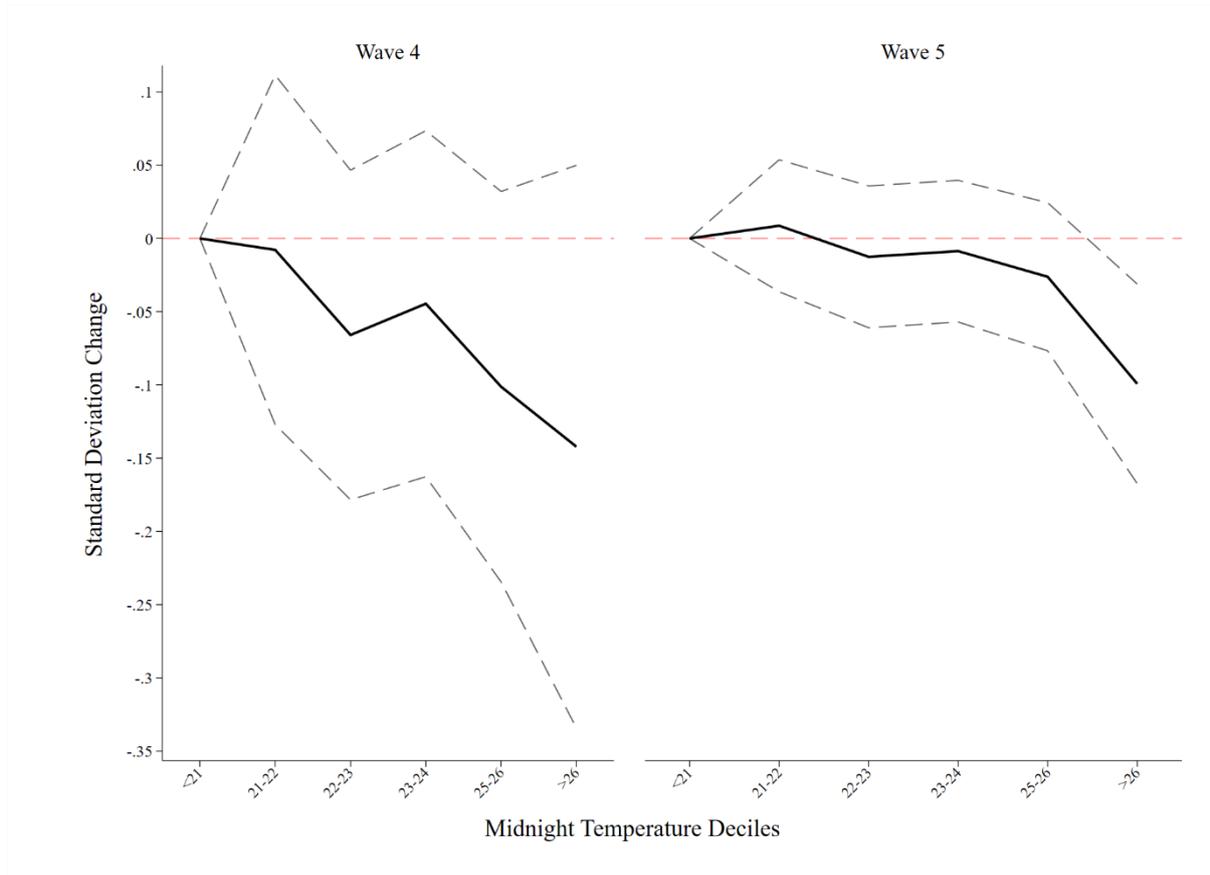

**Notes:** The above figure shows nonlinear effects of midnight temperature on cognition by temperature bins and IFLS waves.

# Appendix B – Additional Results and Robustness Checks

**Appendix Table B1. Correlates of economic preferences and rationality**

|  | Risk Aversion | Rational Choice Violation | Impatience |
|---|---|---|---|
|  | (1) | (2) | (3) |
| Female | 0.072*** | 0.030*** | -0.013** |
|  | (0.007) | (0.006) | (0.006) |
| Age | -0.005*** | -0.001 | 0.008*** |
|  | (0.002) | (0.001) | (0.001) |
| Age$^2$ | 0.000*** | 0.000 | -0.000*** |
|  | (0.000) | (0.000) | (0.000) |
| Religion: Islam | 0.004 | -0.011 | 0.021 |
|  | (0.019) | (0.013) | (0.016) |
| Married or cohabitating | 0.024** | 0.007 | 0.018** |
|  | (0.011) | (0.008) | (0.009) |
| Working in past week | -0.011 | -0.019** | 0.004 |
|  | (0.012) | (0.009) | (0.010) |
| Secondary education or + | 0.012 | -0.056*** | -0.040*** |
|  | (0.009) | (0.007) | (0.007) |
| Cognition global z-score | 0.007 | -0.053*** | -0.055*** |
|  | (0.005) | (0.004) | (0.004) |
| Female head of household | -0.008 | -0.004 | -0.008 |
|  | (0.012) | (0.009) | (0.010) |
| Age of household head | 0.001* | 0.000 | 0.001* |
|  | (0.000) | (0.000) | (0.000) |
| Number of members per household | -0.002 | 0.000 | -0.001 |
|  | (0.002) | (0.002) | (0.002) |
| Household equivalized expenditure | -0.000*** | -0.000*** | -0.000*** |
|  | (0.000) | (0.000) | (0.000) |
| Household equivalized expenditure$^2$ | 0.000*** | 0.000*** | 0.000*** |
|  | (0.000) | (0.000) | (0.000) |
| Outcome Mean | 0.38 | 0.29 | 0.64 |
| Observations | 19,426 | 27,183 | 26,257 |

**Notes:** Models include data from IFLS5. All regressions also include climate zone fixed effects (the result of interacting province of residence, urbanity of the village, altitude groups, and distance to coast groups), plus month and year fixed effects. The regressions exclude individuals living more than 50 kms from the closest grid. Robust standard errors clustered on village level are in parentheses. * p<0.1; ** p<0.05; *** p<0.01.

**Appendix Table B2. Endogenous selection of movers based on preferences**

|  | Δ in Max T (Year of Survey) | Δ in Mean T (Year of Survey) | Δ in Midnight T (Year of Survey) |
|---|---|---|---|
|  | (1) | (2) | (3) |
| Irrational (IFLS4) | -0.016 | -0.051 | -0.034 |
|  | (0.070) | (0.072) | (0.083) |
| Risk Averse (IFLS4) | 0.009 | 0.001 | 0.036 |
|  | (0.072) | (0.072) | (0.083) |
| Impatient (IFLS4) | -0.087 | -0.112 | -0.122 |
|  | (0.067) | (0.068) | (0.078) |
| Age | 0.007 | 0.007* | 0.006 |
|  | (0.005) | (0.004) | (0.004) |
| Female | 0.041 | -0.009 | -0.034 |
|  | (0.031) | (0.031) | (0.036) |
| Married | -0.032 | 0.027 | 0.038 |
|  | (0.070) | (0.071) | (0.080) |
| Religion: Islam | 0.174 | 0.277 | 0.315 |
|  | (0.203) | (0.221) | (0.270) |
| Working in past week | -0.132* | -0.164** | -0.115 |
|  | (0.077) | (0.076) | (0.084) |
| Secondary education or + | -0.023 | -0.004 | -0.019 |
|  | (0.072) | (0.071) | (0.080) |
| Log Expenditure | -0.056 | -0.114** | -0.118* |
|  | (0.053) | (0.053) | (0.062) |
| R-squared | 0.365 | 0.362 | 0.341 |
| Observations | 1,167 | 1,167 | 1,160 |

**Notes:** The change in maximum, average and midnight temperatures (on the year of survey) between the IFLS5 villages - IFLS4 villages for movers only, is regressed on IFLS4 preferences, individual and household characteristics, as well as climate zone, month and year fixed effects. Robust standard errors clustered on village level are in parentheses. * $p<0.1$; ** $p<0.05$; *** $p<0.01$.

**Appendix Table B3. Alternative specifications - Effects of temperature on time and risk preferences**

|  | Rational Choice Violation | | | Impatience | | |
|---|---|---|---|---|---|---|
|  | (1) | (2) | (3) | (4) | (5) | (6) |
| **Panel A** | | | | | | |
| Max Temperature on Day of Survey (°C) | 0.003** | 0.003* | 0.002 | 0.004** | 0.003** | 0.001 |
|  | (0.002) | (0.002) | (0.001) | (0.002) | (0.002) | (0.002) |
| **Panel B** | | | | | | |
| Temperature during hour of survey (°C) | 0.007*** | 0.006*** | 0.004*** | 0.004** | 0.003** | 0.001 |
|  | (0.002) | (0.002) | (0.001) | (0.002) | (0.002) | (0.002) |
| **Panel C** | | | | | | |
| Midnight Temperature Yesterday (°C) | 0.010*** | 0.008*** | 0.007*** | 0.009*** | 0.009*** | 0.006*** |
|  | (0.002) | (0.002) | (0.002) | (0.002) | (0.002) | (0.002) |
| **Panel D** | | | | | | |
| Max Temperature on Day of Survey (°C) | 0.001 | 0.001 | 0.000 | 0.002 | 0.002 | -0.001 |
|  | (0.002) | (0.002) | (0.002) | (0.002) | (0.002) | (0.002) |
| Midnight Temperature Yesterday (°C) | 0.009*** | 0.008*** | 0.007*** | 0.009*** | 0.008*** | 0.007*** |
|  | (0.002) | (0.002) | (0.002) | (0.002) | (0.002) | (0.002) |
| p-value | 0.0180 | 0.0898 | 0.0417 | 0.3335 | 0.1165 | 0.0459 |
| | | | | | | |
| Outcome Mean | 0.31 | 0.31 | 0.31 | 0.62 | 0.62 | 0.62 |
| Observations | 48,953 | 48,953 | 48,953 | 47,566 | 47,566 | 47,566 |
| | | | | | | |
| Humidity | Y | | | Y | | |
| Longitude | | Y | | | Y | |
| Interviewer FE | | | Y | | | Y |

**Notes:** All models include climate zone fixed effects which are the result of interacting province of residence, urbanity of the village, altitude groups, and distance to coast groups. The results exclude individuals living more than 50 kms from the closest grid. All models control for individual and household characteristics and also include linear controls for pollution, precipitation and wind speed. Individual controls include: gender, age, marital status, income generating activity, marital status, religion, highest level of education, day of the week when the individual's survey took place and hour when the survey started. Household controls include: gender of the household head, their age, marital status, education level, number of children and adults in the household, and a squared function of equivalized consumption. Columns 1-6 are estimated on the combined IFLS4 + IFLS5 sample. Robust standard errors clustered on village level are in parentheses. * p<0.1; ** p<0.05; *** p<0.01.

**Appendix Table B4. Robustness check - Sample exclusions and Ramadan**

|  | Rational Choice Violation | | | Impatience | | |
|---|---|---|---|---|---|---|
|  | (1) | (2) | (3) | (4) | (5) | (6) |
| **Panel A** | | | | | | |
| Max Temperature on Day of Survey (°C) | 0.003 | 0.003 | 0.003 | 0.004** | 0.005** | 0.004** |
|  | (0.002) | (0.002) | (0.002) | (0.002) | (0.002) | (0.002) |
| **Panel B** | | | | | | |
| Temperature during hour of survey (°C) | 0.007*** | 0.007*** | 0.007*** | 0.004** | 0.004* | 0.004** |
|  | (0.002) | (0.002) | (0.002) | (0.002) | (0.002) | (0.002) |
| **Panel C** | | | | | | |
| Midnight Temperature Yesterday (°C) | 0.010*** | 0.010*** | 0.009*** | 0.009*** | 0.012*** | 0.010*** |
|  | (0.003) | (0.003) | (0.003) | (0.003) | (0.003) | (0.003) |
| **Panel D** | | | | | | |
| Max Temperature on Day of Survey (°C) | -0.001 | 0.001 | 0.001 | 0.002 | 0.002 | 0.002 |
|  | (0.002) | (0.002) | (0.002) | (0.002) | (0.002) | (0.002) |
| Temperature during hour of survey (°C) | 0.005** | - | - | -0.001 | - | - |
|  | (0.002) |  |  | (0.002) |  |  |
| Midnight Temperature Yesterday (°C) | 0.006* | 0.009*** | 0.009*** | 0.009*** | 0.010*** | 0.009*** |
|  | (0.003) | (0.003) | (0.003) | (0.003) | (0.003) | (0.003) |
| p-value | - | 0.0484 | 0.0990 | - | 0.0730 | 0.0586 |
|  |  |  |  |  |  |  |
| Outcome Mean | 0.31 | 0.31 | 0.33 | 0.62 | 0.62 | 0.62 |
| Observations | 48,953 | 48,953 | 39,067 | 47,566 | 47,566 | 38,038 |
|  |  |  |  |  |  |  |
| Ramadan |  | Y |  |  | Y |  |
| Exclude >1 Visit |  |  | Y |  |  | Y |

**Notes:** Columns 1 and 4 include all three measures of temperature in Panel D, maximum temperature on the day of the survey, temperature at the hour when the survey started, and midnight temperature on the night prior to the survey to assess which is dominant. Columns 2 and 5 include controls for individuals surveyed during Ramadan (1.59% of the sample). Columns 3 and 6 exclude 19% of the sample who received more than 1 visit to complete the time and risk survey modules. The p-values in panel D test whether the coefficients of maximum and midnight temperature are significantly different from each other. All models include climate zone fixed effects (the interaction of province-urban-altitude-distance to coast), plus month and year fixed effects and exclude individuals living more than 50 kms from the closest grid. Models also include village altitude, environmental, individual and household level controls. All models are estimated on the combined IFLS4 + IFLS5 sample. Robust standard errors clustered on village level are in parentheses.

* $p<0.1$; ** $p<0.05$; *** $p<0.01$.

**Appendix Table B5. Only climate zones where date of survey explains >90% of the variation in temperatures**

|  | Most Risk Averse | Rational Choice Violation | Most Impatient |
|---|---|---|---|
|  | (1) | (2) | (3) |
| **Panel A** | | | |
| Max Temperature on Day of Survey (ºC) | 0.000 | 0.002 | 0.005** |
|  | (0.002) | (0.002) | (0.002) |
| **Panel B** | | | |
| Temperature during hour of survey (ºC) | 0.001 | 0.005** | 0.004** |
|  | (0.002) | (0.002) | (0.002) |
| **Panel C** | | | |
| Midnight Temperature Yesterday (ºC) | 0.002 | 0.008*** | 0.010*** |
|  | (0.003) | (0.003) | (0.003) |
| **Panel D** | | | |
| Max Temperature on Day of Survey (ºC) | -0.000 | -0.001 | 0.002 |
|  | (0.002) | (0.002) | (0.002) |
| Midnight Temperature Yesterday (ºC) | 0.002 | 0.009*** | 0.009*** |
|  | (0.003) | (0.003) | (0.003) |
| p-value | 0.6819 | 0.0458 | 0.1009 |
|  |  |  |  |
| Outcome Mean | 0.37 | 0.32 | 0.62 |
| Observations | 28,348 | 43,452 | 42,227 |

Notes: All models include climate zone fixed effects which are the result of interacting province of residence, urbanity of the village, altitude groups, and distance to coast groups. The results exclude individuals living more than 50 kms from the closest grid and those who live in climate zones where date of survey fixed effects explain less than 90% of the variation in temperatures. All models control for individual and household characteristics. Individual controls include: gender, age, marital status, income generating activity, marital status, religion, highest level of education, day of the week when the individual's survey took place and hour when the survey started. Household controls include: gender of the household head, their age, marital status, education level, number of children and adults in the household, and a squared function of equivalized consumption. Columns 1-3 are estimated on the combined IFLS4 + IFLS5 sample. Robust standard errors are clustered on village level in parentheses. * $p<0.1$; ** $p<0.05$; *** $p<0.01$

**Appendix Table B6. Placebo test - Maximum temperature in 2 weeks**

|  | Rational Choice Violation | Impatience |
|---|---|---|
|  | (1) | (2) |
| **Panel A** |  |  |
| Max Temperature on Day of Survey (ºC) | 0.003 | 0.003* |
|  | (0.002) | (0.002) |
| Max Temperature on Day T+14 (ºC) | 0.002 | 0.002 |
|  | (0.002) | (0.002) |
| **Panel B** |  |  |
| Temperature during hour of survey (ºC) | 0.007*** | 0.003* |
|  | (0.002) | (0.002) |
| Max Temperature on Day T+14 (ºC) | 0.001 | 0.002 |
|  | (0.002) | (0.002) |
| **Panel C** |  |  |
| Midnight Temperature Yesterday (ºC) | 0.009*** | 0.009*** |
|  | (0.003) | (0.003) |
| Max Temperature on Day T+14 (ºC) | 0.001 | 0.001 |
|  | (0.002) | (0.002) |
| **Panel D** |  |  |
| Max Temperature on Day of Survey (ºC) | 0.000 | 0.001 |
|  | (0.002) | (0.002) |
| Midnight Temperature Yesterday (ºC) | 0.009*** | 0.008*** |
|  | (0.003) | (0.003) |
| Max Temperature on Day T+14 (ºC) | 0.001 | 0.001 |
|  | (0.002) | (0.002) |
| Outcome Mean | 0.31 | 0.62 |
| Observations | 48,953 | 47,566 |

**Notes:** All models include climate zone fixed effects, the interaction of province-urban-altitude groups-distance to coast groups, plus month and year fixed effects. All regressions exclude individuals living more than 50 kms from the closest grid. Models also include village altitude, environmental, individual and household level controls. Robust standard errors clustered on village level are in parentheses. * $p<0.1$; ** $p<0.05$; *** $p<0.01$.

**Appendix Table B7. Placebo Test - Temperature and Demographics**

|  | Female | Age | Married | Equivalised Expenditure |
|---|---|---|---|---|
|  | (1) | (2) | (3) | (4) |
| **Panel A** | | | | |
| Midnight Temperature Yesterday (°C) | 0.000 | 0.004 | 0.001 | -1.0e+04 |
|  | (0.002) | (0.091) | (0.003) | (1.0e+04) |
| **Panel B** | | | | |
| Max Temperature on Day of Survey (°C) | 0.000 | 0.020 | 0.001 | 8561.983 |
|  | (0.001) | (0.059) | (0.002) | (6110.675) |
| **Panel C** | | | | |
| Mean Temperature on Day of Survey (°C) | -0.001 | -0.030 | 0.001 | 862.665 |
|  | (0.002) | (0.091) | (0.003) | (9472.537) |
| Observations | 54,102 | 53,598 | 53,604 | 53,570 |

**Notes:** All models include climate zone fixed effects which are the result of interacting province of residence, urbanity of the village, altitude groups, and distance to coast groups. Columns 1-4 are estimated on the combined IFLS4 + IFLS5 sample. Robust standard errors are clustered on village level in parentheses. * $p<0.1$; ** $p<0.05$; *** $p<0.01$

**Appendix Table B8. Potential mechanisms behind temperature effects - Cognition**

| | Midnight Temperature Yesterday (ºC) | | |
|---|---|---|---|
| Outcomes | IFLS4 age 15-31 | IFLS5 age 15-31 | IFLS5 full sample |
| | (1) | (2) | (3) |
| Cognition global z-score | -0.036** | -0.005 | -0.016** |
| | (0.017) | (0.008) | (0.006) |
| TICS z-score | -0.033** | 0.005 | -0.004 |
| | (0.015) | (0.009) | (0.006) |
| Raven's Matrix z-score | -0.006 | -0.015** | -0.020*** |
| | (0.013) | (0.007) | (0.005) |
| Math z-score | -0.039*** | -0.001 | -0.013** |
| | (0.014) | (0.009) | (0.006) |
| Observations | 7,131 | 11,026 | 26,900 |
| Bin FE | Y | Y | Y |
| Month FE | Y | Y | Y |
| Year FE | Y | Y | Y |
| Latitude | Y | Y | Y |
| Pollution, Rain & Wind | Y | Y | Y |

**Notes:** Models include controls for village altitude and environmental variables such as pollution, precipitation and wind, and exclude individuals living more than 50 kms from the closest grid. Robust standard errors clustered on village level are in parentheses. * $p<0.1$; ** $p<0.05$; *** $p<0.01$.

**Appendix C – Cognition Outcomes**

We measure the effects of temperature on four different tests of cognition and mathematical numeracy. We separate this analysis by IFLS wave as there were between-wave differences in the composition of these tests. Specifically, IFLS5 asked these four tests of all individuals age 15-90 years old (who also played the lottery games). However, in IFLS4, some of these tests were only asked of adults age 15-31 and one item, fluid intelligence, was not included at all.

The first test we examine, *Fluid Intelligence,* is an adaptive-number series test, which "measures a specific form of fluid intelligence related to quantitative reasoning" (Strauss et al., 2017, p. 6). It is based on the test used in the Health and Retirement Study (Fisher et al., 2014) and found to be closely correlated with financial wealth (Smith et al., 2010). It consists of a series of numbers where one of the values is missing and the respondent must fill in the missing number. The difficulty of the subsequent series is increasing with the number of correct answers. These are then scored using a standardized score called a W-Score, which appears as *Fluid Intelligence z-score* in Table 1.

The second test we examine is designed after the Telephone Interview of Cognitive Status (Fong, et al., 2009), and measures mental intactness. This test includes a date awareness and word recall module. In IFLS5, it also includes a set of subtraction questions that ask the respondent to subtract 7 from 100, five successive times, and a final exercise where the respondent draws a figure of two overlapping pentagons. The number of correct answers is added to generate a score from 0 – 30, this score is then standardized to create what we call in Table 1 the *TICS z-score*.

We also use the Raven's Progressive Matrices, which measures non-verbal abilities (Raven, 2000), and a test of mathematical skills which includes a series of addition, subtraction, division and multiplication questions. These two tests were asked of all adults inIFLS5, but only adults 15-31 years old in IFLS4. For both tests, we calculate the percentage of correct answers and then standardize the resulting scores to create the *Raven's Progressive Matrices z-score* and *Math z-score*. Finally, we create a composite measure of cognition by performing principal component analysis with the z-scores of all the items described above: fluid intelligence z-score, TICS z-score, Raven's Matrices, and math z-scores. The resulting global score is then standardized.